\documentclass[journal, onecolumn, 10pt]{IEEEtran} 
\usepackage{amsfonts}
\usepackage{graphics}
\usepackage{latexsym}
\usepackage{amsmath,amsthm}
\usepackage{dsfont}
\usepackage[dvips]{epsfig}
\usepackage{algorithm, algorithmic}
\usepackage{url}
\usepackage{tikz}

\usepackage{soul}

\usepackage[numbers,sort&compress]{natbib}

\newtheorem{Theorem}{Theorem}

\newtheorem{Example}{Example}
\newtheorem{Remark}{Remark}
\def\bit{\bibitem}

\def\IM{{\textit{I}-Measure}}

\def\Xitip{Xitip}
\def\st{Shannon-type}
\def\nst{non-Shannon-type}

\begin{document}
\title{A Numerical Study on the Wiretap Network with a Simple Network Topology}
\author{
Fan Cheng,~\IEEEmembership{Member,~IEEE} and  Vincent Y. F. Tan,~\IEEEmembership{Senior Member,~IEEE}
\thanks{F.\ Cheng  is with the Department of Electrical and Computer Engineering, National University of Singapore, Singapore. Email: fcheng@nus.edu.sg.}
\thanks{V.~Y.~F.\  Tan  is with the Department of Electrical and Computer Engineering and the Department of Mathematics, National University of Singapore, Singapore. Email: vtan@nus.edu.sg.}
 \thanks{This paper was presented in part at the International Symposium on Information Theory, 2014 \cite{cheng14isit}.}
}

\maketitle

\begin{abstract}
In this paper, we study a security problem on a simple wiretap network, consisting of  a source node S, a destination node D, and an intermediate node R. The intermediate node connects the source and the destination nodes via a set of noiseless parallel channels, with sizes $n_1$ and $n_2$, respectively.  A message $M$ is to be sent from S to D.
The information in the network may be eavesdropped by a set of wiretappers. The wiretappers cannot communicate with one another. Each wiretapper can access a subset of channels, called a wiretap set.   All the chosen wiretap sets form a wiretap pattern.  A random key $K$ is generated at S and a coding scheme on $(M, K)$ is employed  to protect  $M$.  We define two decoding classes at  D: In Class-I, only $M$ is required  to be recovered and in Class-II, both  $M$ and  $K$ are required to be recovered. The objective is to minimize $H(K)/H(M)$ {for a given wiretap pattern} under the  perfect secrecy constraint. The first question we address is whether routing is optimal on this simple network. By  enumerating all the wiretap patterns on the Class-I/II $(3,3)$ networks and harnessing the power of \st\ inequalities, we find that gaps exist between the bounds implied by routing and the bounds implied  by \st\ inequalities for a small fraction~($<2\%$) of all the wiretap patterns. The second question we investigate is the following: What is  $\min H(K)/H(M)$ for the remaining wiretap patterns where gaps exist?
We study some simple wiretap patterns and find that their  Shannon bounds (i.e., the lower bound induced by \st\ inequalities)  can be achieved by linear codes, which means routing is not sufficient even for the ($3$, $3$) network.  For some complicated wiretap patterns, we study the structures of  linear coding schemes under the assumption that they can achieve the corresponding Shannon bounds. The study indicates that
the determination of the entropic region of $6$ linear vector spaces  cannot be sidestepped.
Some subtle issues on the network models are  discussed and  interesting observations are stated.
\end{abstract}


\section{Introduction}
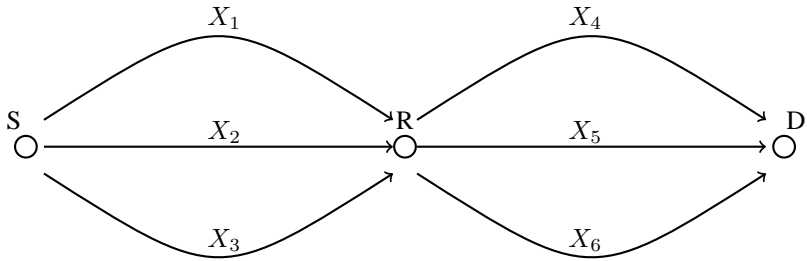
\begin{figure}[!htb]
  \centering
  \begin{tikzpicture}[xscale=0.8,yscale=0.8]
\draw  [ thick](-0.3,1.854) circle [radius=0.18];
\draw  [ thick](6,1.854) circle [radius=0.18];
\draw  [ thick](12+0.3,1.854) circle [radius=0.18];

\draw [thick, ->] (0,1.854*2-2.3) .. controls (2.9 ,1.854-2.3)  .. (6-0.2, 1.854*2-2.3);
\draw [thick, ->] (6+0.2, 1.854*2-2.3) .. controls (9.1 ,1.854-2.3)  .. (12, 1.854*2-2.3);

\draw [ thick, ->] (0,1.854) -- (6-0.2,1.854);
\draw [ thick, ->] (6+0.2, 1.854) -- (12, 1.854);

\draw [thick, ->] (0,2.3) .. controls (2.9 ,1.854+2.3)  .. (6-0.2, 2.3);
\draw [thick, ->] (6+0.2, 2.3) .. controls (9.1 ,1.854+2.3)  .. (12, 2.3);

\node at (-0.5, 2.3) {S};
\node at (6, 2.3) {R};
\node at (12+0.5, 2.3) {D};

\node at (1.5*2, 4) {$X_{1}$};
\node at (1.5*2, 2.1) {$X_{2}$};
\node at (1.5*2, 0.3) {$X_{3}$};
\node at (4.5*2, 4) {$X_{4}$};
\node at (4.5*2, 2.1) {$X_{5}$};
\node at (4.5*2, 0.3) {$X_{6}$};
\end{tikzpicture}
  \caption{The communication model with three nodes and six noiseless channels}\label{general-model-33}
\end{figure}

\subsection{A Security Problem on a Simple Communication Network }

In this paper, we study  a security problem on a communication network (depicted in Fig. \ref{general-model-33}) with three nodes S, R, and D, where S is the source node, R is the intermediate node, and D is the destination node, respectively. There are three noiseless channels connecting the pairs (S, R) and (R, D).

A private message $M$ is generated at S and is to be sent to D. As there is a collection of wiretappers that  can only tap the information on a subset of the channels,  a random key $K$ which is independent of $M$ is also generated at node S.   {To protect the message $M$, a coding scheme whose encoders take as inputs both $M$ and $K$ is employed to combat the effect of the wiretappers.} This coding scheme ensures that the information read by each wiretapper is independent of $M$. Furthermore, wiretappers cannot communicate with one another. For each wiretapper, the set of channels it accesses, called a {\em wiretap set}, is fixed before the coding scheme  on $(M,K)$ is constructed. The set of all the wiretap sets is called a {\em wiretap pattern}. Now, a fundamental question arises: If we fix the size of the message, what is the minimum size of the key should be injected to protect the message? Here the ``size'' of a random variable is measured by its entropy. The problem  we stated does not  seem  to be very hard since the network topology is simple and a simpler version of this problem dates back to Shannon~\cite{shannon1998communication}.  However, as we shall show in this paper, this problem appears to be fiendishly hard. Even for such a simple network, the problem is challenging to solve completely.

\subsection{Related Results}
\subsubsection{Network Coding}
We leverage two   important concepts in this paper, namely,  {\em routing} and {\em network coding}.  In most communication networks, information is transmitted in a \textit{store-and-forward} manner; i.e., bits  are delivered as commodities  and then are   routed from a node to another. The bits are unaltered on the transmission paths.
Ahlswede \textit{et~al.}~\cite{ahlswede2000network} proposed a
network communication  paradigm called \textit{network coding}, where the role of the intermediate nodes is enhanced as follows.
At each intermediate node, the information received on the input
channels may be encoded, and may be
sent on the output channels.  Network coding can increase the achievable rates and even attain the capacity of the network. In~\cite{ahlswede2000network}, the classical \rm{max-flow-min-cut} theorem is generalized to multicast scenario.
{Furthermore, the authors demonstrated that network coding can outperform \textit{store-and-forward} in terms of bandwidth utilization.}
Routing is a class of special network coding schemes. Indeed, when network coding is used, the information is coded in the  network.
We refer the reader to Yeung~\textit{et al.}~\cite{yeung2006network}
for a comprehensive treatment of network coding theory.

\subsubsection{Information-Theoretic  Security and Wiretap Networks}
Information-theoretic security was launched in Shannon's  seminal work \cite{shannon1998communication}, where the communication model is only a single channel. A key $K$ is stored at the  sender and receiver before the  message $M$ is sent. The sender generates $X$ from $M$ and $K$ by an encoding function. Then $X$ is sent through the channel. The receiver decodes $M$ from $X$ and $K$. The main result, called the {\em perfect secrecy theorem}, implies that the  size of $K$ (measured by its entropy) is lower bounded by the  size of $M$.

In   wiretap networks, Cai and Yeung \cite{cai2002secure}
considered sending a private message to possibly more than one receiver
through a noiseless communication network. Their model is as follows:
The communication may be eavesdropped by a set of wiretappers, who cannot communicate with one another. Each of the wiretappers
can access a subset of channels of the network, called a {\em wiretap set}.
The wiretapper can choose an arbitrary wiretap set before communication and the choice is fixed during communication.
The set of all the chosen wiretap sets is commonly referred to as a  {\em   wiretap pattern}, denoted by $\mathcal{A}$.
For such an $\mathcal{A}$, the sender and the legitimate users have to design a coding scheme to combat the effect of the wiretappers.
The strategy is to generate a random key $K$ to protect the message and  send both the  message and the key via a network coding scheme. This ensures  that a
wiretapper can only observe some functions of the message and the
key, where the output of the functions are statistically independent of the message.
On the other hand, a receiver node can recover the private message
by decoding the information received from its input channels. The performance of a secure network coding scheme is measured
by   the sizes of the message and the key.
In designing a secure network coding scheme, the aim is to maximize
the size of the message whist minimizing the size of the
key. In \cite{cai2002secure}, when  $\mathcal{A}$
consists of all subsets of channels whose sizes are at
most some constant $r$,  there exists a linear network code which
is optimal in both the sizes of the message and the key. When $\mathcal{A}$ is arbitrary, a cut-set bound on the  ratio of the size of the key and the size of the message was obtained in Cheng~and~Yeung~\cite{ChengYeung2011NetC, CY2014IT}.
The wiretap network model is a generalization of the well known perfect secrecy system studied by Shannon \cite{shannon1998communication}, and
is also widely studied from many different perspectives. A  comprehensive survey of the fundamental performance limits in wiretap networks can be found in Cai~and~Chan~\cite{Cai-Chan-SNC11}. 

From the point of view of equivalence and complexity, Cui \textit{et al.} \cite{CHK13} showed that determining the secrecy capacity is an NP-hard problem for nonuniform or restricted wiretap sets. Huang \textit{et al.} \cite{HTLK13}  studied the problem where  network nodes can independently generate randomness and  the wiretap sets are uniform, showing that determining the secrecy capacity is at least as difficult as the $k$-unicast network coding problem. More results on equivalence of network coding problems can be found in Huang \textit{et al.} \cite{HLJ15}. {We emphasize that the focus of this paper is {\em not} on determining the secrecy capacity; rather we are concerned with finding the minimum key size relative to the message length, i.e., $H(K)/H(M)$. Thus, we avoid the difficulties pointed out in \cite{CHK13, HTLK13}; yet we encounter a different set of difficulties.}

\subsubsection{Shannon- and \nst\ Inequalities}

The properties of Shannon information measures form a useful set of tools to investigate the properties of wiretap networks.
For a set of random variables, the  properties of information measures such as entropy, mutual information, conditional entropy are well known.
In particular, it is well known that the above information measures are non-negative.
Information  inequalities which can be implied by Shannon's
information measures are referred to as  \st\
inequalities; e.g., the inequality $H(X_1|X_2) \leq H(X_1)$. If only the \st\ inequalities are concerned,
a one-to-one correspondence between information measures and set
theory can be established in the so-called \IM\  theory by Yeung~\cite[Ch. 3]{yeung2008information}, which only involves  simple set theory operations; i.e., union, intersection, complement, and set difference.
Moreover, the following fundamental result~\cite[Ch.\ 14]{yeung2008information}  was established for the information measures: \\
Let $[n]=\{1, 2, \ldots, n\}$. Any Shannon's information measures involving random variables $X_1$, $X_2$, $\ldots$,  $X_n$ can be expressed as the sum of the following two elemental forms:
\begin{itemize}
\item [(i)]$ H(X_i|X_{[n]-\{i\}}), i\in [n]$;
\item [(ii)] $I(X_i;X_j|X_{K})$, where $i\neq j$ and $K\subseteq [n]-\{i,j\}$.
\end{itemize}
Note that all the information inequalities we studied are {\em linear}.
If we regard these $n+\binom{n}{2}2^{n-2}$ elements as variables, then any information expression can be rewritten as a linear combination of them.
This observation enables us to check the correctness of  a \st\ inequality by transforming it into an equivalent linear program, which can be easily implemented in a computer program. ITIP~\cite{ITIP} and \Xitip\ \cite{Xitip} are two widely used software packages based on this very principle, where the latter is an upgraded
version of the former. When a certain information inequality  and some constraints are supplied as inputs to the software program, it will inform the user whether the information inequality is a valid \st\ inequality.
The latest extensions of ITIP may be found in Tian~\cite{Tian433} and  Ho~\textit{et al.}~\cite{HoTanYeung14}.

It was a long-standing problem as to whether there exist
inequalities involving information measures that cannot be directly implied by  \st\
inequalities. This was before the seminal work by Zhang~and~Yeung~\cite{ZY98}, where the first \nst\ inequality  was
proved. All the information inequalities on $n$ random variables characterize the so-called entropic region $\bar{\Gamma}_n^{*}$ \cite{yeung2008information}. Note that the \IM-based method is futile for proving 
non-Shannon-type inequalities. In Dougherty \textit{et al.}
\cite{DouZeger2007}, the inequality in \cite{ZY98} was used to
reduce the capacity bound in a communication network, indicating
that \st\ inequalities are not always sufficient in practice.  The
general theory of \nst\ inequalities  is still in its
infancy and relatively little progress has been achieved. So far, the problem has
been addressed when the number of random variables $n$~$\leq$~$3$. When all the random variables are in one-to-one correspondences with vector spaces, the problem has been settled for $n\leq 5$ in Dougherty~\textit{et al.}~\cite{DFZ2010}.
For the case $n = 6$, Dougherty~\cite{Dougherty2014} showed that the number of different linear rank inequalities  exceeds $1$ million.  The exact set of linear rank inequalities is still unknown.

\subsubsection{Network Coding meets Information Inequalities}
In essence, network coding problems can be perfectly represented by the aforementioned information measures.
We can use random variables to denote the information transmitted on the channels.
The encoding and decoding process at each node can be dealt by information equalities. The performance bounds of network coding can be expressed via information inequalities.
It was proved in Chan~and~Grant~\cite{ChanGrant2012} that the general secure network coding problem on multi-source and multi-sink network is as hard as determining the exact
entropic region. That is, the general secure network coding problem is hard to solve. Thus, we seek bounds. We may consider only \st\ inequalities to obtain a bound on
a concrete problem by solving the corresponding linear program, or
invoking some \IM-based softwares; e.g., ITIP or \Xitip. In the sequel, we refer to the bound obtained by \st\ inequalities  as the \textit{Shannon bound}.
However, the \IM-based method suffers from the  drawback that the
computational complexity is exponential in  the number of random
variables, which means it works well only for very few random
variables. Fortunately, we may also consider  the cut-set bound, which is a  classic tool in analyzing
the performance of network coding. In some situations, the cut-set bound is tight;
e.g.,  the bound obtained in \cite{cai2002secure}.

\subsection{The Problem We Study and The Question We Ask}
Network coding trumps routing in many aspects of communication scenarios. However, routing is advantageous over network coding due to its lower complexity in encoding and decoding and it is easy to understand and analyze. In some simple networks, in lieu of network coding, routing can be shown to be sufficient. In the wiretap network, the only easily-computable bounds are the cut-set bound and Shannon bound. {The cut-set bound is optimal for the point-to-point communication system.}  If we consider only routing in the network, we can transform the general network model  to a point-to-point network, which means we can obtain  a  bound  based on routing. Here we refer to it as the \textit{routing bound}.  For a point-to-point communication network, all these three bounds are tight; while for general networks, none of them is tight in general.  The main motivation of this work is from the following fundamental question:

\begin{quote}
 {\em Can we systematically assess the tightness of these bounds in a marginally more  complex  network compared to a point-to-point network?}
\end{quote}

\subsection{Main Contribution and Techniques}

In this paper, we assess the optimality of routing in the wiretap network with a simple topology (Fig.\ \ref{general-model-33}). Recall that in~\cite{ChengYeung2011NetC, CY2014IT}, the cut-set bound is tight for a point-to-point communication system.   Beyond the cut-set bound, no further result has been known till date and  it is not clear to what extent, the routing scheme is optimal. Our network model is more complex since there is an extra encoding node. It is interesting to deduce whether routing is optimal.  In this paper, we use ITIP/Xitip to compare  the routing bound,  the cut-set bound, and the Shannon bound for different wiretap patterns. We find out that for some wiretap patterns, there are  gaps between the  routing bounds  and  the Shannon bounds. We pick some examples from these wiretap patterns with gaps, and construct  coding schemes to achieve the Shannon bounds. Hence, one of the main takeaways is that the routing bound is not tight even in this simple network. For some concrete wiretap patterns where gaps exist, we discuss linear coding schemes to achieve the Shannon bounds. Furthermore, two different decoding classes  are defined and  the distinction between them is discussed.

We summarize our key contributions in this paper:  Firstly, we enumerate all the possible wiretap patterns and determine the wiretap patterns where gaps between Shannon bounds and routing bounds exist. Secondly, for some wiretap patterns, we construct coding schemes to show that routing is not optimal. Lastly, we study some complex wiretap patterns and present linear coding scheme for these patterns. By doing so, we gain an intuitive  understanding of why the Shannon bound is, in general,  difficult to attain in this simple network. We make several interesting observations during the course of our numerical study. We also discuss some interesting problems inspired by the study.

\subsection{Paper Organization}
The paper is organized as follows. In Section \ref{sec1}, the problem formulation is stated and some related results are discussed. In Section \ref{sec2}, we explain how the experiments are conducted and discuss the results. From Section \ref{sec3} to Section \ref{subop32}, we select some wiretap patterns to further validate our claims. In Section \ref{hdexam}, we study two hard wiretap patterns and the structures of linear coding schemes to achieve  the  Shannon bounds. We conclude the paper in Section \ref{sec4} by summarizing our key contributions and stating directions for further work.

\section{Problem formulation}\label{sec1} 
The general problem formulation is described as follows (depicted in Fig. \ref{general-model}):
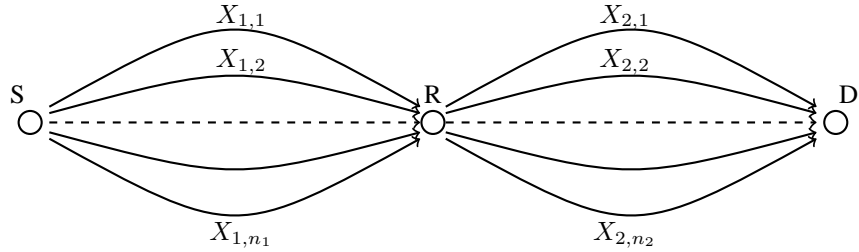
\begin{figure}[!htp]
  \centering
  \begin{tikzpicture}[xscale=0.85,yscale=0.85]

\draw  [ thick](-0.3,1.854) circle [radius=0.18];
\draw  [ thick](6,1.854) circle [radius=0.18];
\draw  [ thick](12+0.3,1.854) circle [radius=0.18];

\draw [ thick, ->] (0,1.6) .. controls (2.9, 0) ..  (6-0.2,1.6);
\draw [ thick, ->] (6+0.2, 1.6) .. controls (9.1, 0) ..(12, 1.6);

\draw [ thick, ->] (0,1.7) .. controls (2.9, 0.927) .. (6-0.2,1.7);
\draw [ thick, ->] (6+0.2, 1.7) .. controls (9.1, 0.927) .. (12, 1.7);

\draw [ thick, dashed, ->] (0,1.854) -- (6-0.2,1.854);
\draw [ thick, dashed, ->] (6+0.2, 1.854) -- (12, 1.854);

\draw [ thick, ->] (0,2)  .. controls (2.9, 0.927*3) .. (6-0.2,2);
\draw [ thick, ->] (6+0.2, 2) .. controls (9.1, 0.927*3) .. (12, 2);

\draw [ thick, ->] (0,2.1) .. controls (2.9, 0.927*4) ..(6-0.2, 2.1);
\draw [ thick, ->] (6+0.2, 2.1) .. controls (9.1, 0.927*4) .. (12, 2.1);

\node at (-0.5, 2.3) {S};
\node at (6, 2.3) {R};
\node at (12+0.5, 2.3) {D};

\node at (1.5*2, 3.5) {$X_{1,1}$};
\node at (1.5*2, 2.8) {$X_{1,2}$};
\node at (1.5*2, 0.1) {$X_{1, n_1}$};

\node at (4.5*2, 3.5) {$X_{2,1}$};
\node at (4.5*2, 2.8) {$X_{2,2}$};
\node at (4.5*2, 0.1) {$X_{2,n_2}$};
\end{tikzpicture}
  \caption{The general communication model}\label{general-model}
\end{figure}

\begin{itemize}
\item [1.] The network consists of three nodes, the source node S, the intermediate node  R, and the destination node D. There are  noiseless directed edges (channels) connecting the pairs (S, R) and (R, D). Denote the set of channels by $\mathcal{E}$. Let $n_1$ be the number of channels from S to R and $n_2$ be the number of channels from R to D, respectively.
To simplify our discussion, we assume that the capacity of each channel is much larger than the sum of information rates.
\item [2.]  {Let $\mathcal{M}$ and $\mathcal{K}$ be the finite alphabets of  the message $M$ and the key $K$, respectively.} At the source node S, a pair of uniformly distributed message and private key $(M, K)$ is generated, where $M\in \mathcal{M}$ is the message and $K\in \mathcal{K}$ is the private key. It is assumed that $M$ and $K$ are statistically independent; i.e.,
    \begin{equation}
        I(M;K) = 0.
    \end{equation}
\item [3.] Denote the information transmitted on the channels from S to R by $X_{1,1}, X_{1,2},..., X_{1, n_1}$. Then
\begin{equation}
    H(X_{1,1}, X_{1,2},...,X_{1,n_1}|M,K) = 0.
\end{equation}
 All the above $X_{1,i}$'s belong to the finite alphabets $\mathcal{X}_{1,i}$'s for $1\leq i\leq n_1$.
\item [4.] Denote the information transmitted on the  channels from R to D by $X_{2,1}, X_{2,2},..., X_{2, n_2}$. Then
\begin{equation}
    H(X_{2,1}, X_{2,2},...,X_{2,n_2}|X_{1,1}, X_{1,2},...,X_{1,n_1}) = 0.
\end{equation}
 All the above $X_{2,i}$'s belong to the finite alphabets $\mathcal{X}_{2,i}$'s for $1\leq i\leq n_2$. {Note that when one designs a code for the network, s/he is allowed to not only design the random variables $\{ (X_{1,i}, X_{2,j}) :i = 1,\ldots, n_1, j = 1,\ldots, n_2 \}$ but also their finite alphabets $\{ (\mathcal{X}_{1,i},\mathcal{X}_{2,j}) :i = 1, \ldots, n_1,j = 1,\ldots, n_2 \}$.}
\item [5.] At the destination node D, we set two different decoding classes. The first is only  $M$ is needed to be decoded at D and the other is that both $M$ and $K$ are needed to be decoded at D.
We refer to these two different models as Class-I $(n_1,n_2)$ network and Class-II $(n_1,n_2)$ network, respectively. Specifically, in Class-I, we have
\begin{equation}
H(M|X_{2,1}, X_{2,2},...,X_{2,n_2}) = 0;
\end{equation}
and in Class-II, we have
\begin{equation}
H(M,K|X_{2,1}, X_{2,2},...,X_{2,n_2}) = 0.
\end{equation}
\item [6.] There is a set of wiretappers, each of which can access an arbitrary subset of $\mathcal{E}$. The choice of the set each wiretapper selects is fixed before communication commences and stays the same during the process of communication. Wiretappers cannot communicate with one another.
    The set of  choices of the wiretappers is denote by  $\mathcal{A} \subseteq 2^{\mathcal{E}}$. In the sequel,  $\mathcal{A}$ is referred to as a wiretap pattern. The sender and receiver need to consider all $A\in \mathcal{A}$, simultaneously. For a set $A\in \mathcal{A}$, denote $(X_e, e\in A)$ by $X_A$. In this model, perfect secrecy is required. To be concrete,
    \begin{equation}\label{constraint:sec}
        I(M;X_{A}) = 0,\quad \forall A \in \mathcal{A}.
    \end{equation}
\item [7.]
In this work, we are interested in minimizing $H(K)/H(M)$, given the constraints above; i.e., we are interested in finding
\begin{equation}
\min \frac{H(K)}{H(M)}, \text{ for a given $\mathcal{A}$}. \footnote{ {In principle, ``$\min$'' should be replaced by ``$\inf$'', since in general, the lower bound on ${H(K)}/{H(M)}$ can be only attained  by random variables with countably infinite alphabets, so the constraint set is not closed. For ease of expositions, we use ``$\min$'' here and in the following.}}
\end{equation}
Specifically, if we fix the size of $M$, what is the minimal size of $K$  to achieve perfect secrecy?
\end{itemize}
The encoding and decoding functions at S, R, and D should abide by the conventions in network coding theory \cite{yeung2006network}. Moreover, the secrecy constraints~(\ref{constraint:sec}) should be satisfied. The encoder  at the sender S is
\begin{equation}
f_S: \mathcal{M}\times\mathcal{K}\to \prod_{i=1}^{n_1}\mathcal{X}_{1,i}.
\end{equation}
The encoder at node R is
\begin{equation}
f_R: \prod_{i=1}^{n_1}\mathcal{X}_{1,i} \to  \prod_{i=1}^{n_2}\mathcal{X}_{2,i}.
\end{equation}
The decoder at node  D is:
\begin{equation}
f_{\rm{I}}: \prod_{i=1}^{n_2}\mathcal{X}_{2, i} \to \mathcal{M};\qquad \text{ (Class-I)}
\end{equation}
or
\begin{equation}
f_{\rm{II}}: \prod_{i=1}^{n_2}\mathcal{X}_{2, i} \to \mathcal{M}\times\mathcal{K}. \qquad \text{ (Class-II)}
\end{equation}


The Class-I model is a special case of the wiretap network introduced in \cite{cai2002secure}.  In Class-I networks, the private key $K$ may be operated on  at the intermediate node R to potentially increase the message size whilst ensuring that the message is transmissible securely over the network. {The  Class-II model is applicable to the scenario where multiple messages are to be delivered to the receivers and some  non-important messages are used to protect the remaining important ones.} In our model, since we have removed the capacity constraints
on the channels (capacities are assumed to be sufficiently large), it is interesting to understand the impact of the condition that $K$ must be recovered at the destination. In particular, we ask whether there are  any wiretap patterns for which  $\min H(K)/H(M)$ is changed when the decoding requirements on $K$ are different. The answer is, in general, yes.

\subsection{Cut-set bound,  Routing bound, Shannon bound}
First, we state the cut-set bound for an arbitrary $\mathcal{A}$.
\begin{Theorem}[\cite{ChengYeung2011NetC,CY2014IT}]
Let $W=\{e_1,e_2,...,e_n\}$ be a cut-set and $\mathcal{A}\subseteq 2^{W}$ be a wiretap pattern. Then
\begin{equation}
    \frac{ H(K)}{ H(M) } \geq \frac{1}{\max\sum\limits_{i=1}^{n}x_i-1},
\end{equation}
where
\begin{equation}\label{cut-set-constraint}
    \sum_{e_i\in A} x_i \leq 1, \text{ $\forall$ $A\in\mathcal{A}$,}
\end{equation}
and
\begin{equation}
    x_i\geq 0,\ \ 1\leq i\leq n.
\end{equation}
\end{Theorem}
This cut-set bound can be interpreted as follows. Since the focus is on the ratio between $H(K)$ and $H(M)$,  we may as well set $H(K)$ to be $1$.  Let $x_i$ be the information rate on channel (represented by edge) $e_i$, $1\leq i \leq n$. Assume that the symbols on the channels are mutually independent, then the constraints (\ref{cut-set-constraint}) mean that the size of the symbols in each wiretap set cannot exceed the size of the key. Furthermore, the cut-set bound is tight for a point-to-point network and  its optimality can be achieved by a linear code. Hence, if we wish to know to what extent  routing is optimal, we should consider a network with at least three nodes. {The algorithm for computing the cut-set bound is described in  Alg. 1.}
\begin{algorithm}\label{algorithm:1}
\caption{Algorithm for computing the cut-set bound}
\begin{itemize}
 \item [1] Denote the number of edges in the cut-set by $n$ and the number of wiretap sets by $d$. Let $A$ be a $d \times n$ matrix.
 \item [2] If the $i$th wiretapper can access the $j$th edge, then $A(i, j) = 1$. Otherwise, $A(i,j) = 0$.
 \item [3] Let $1_d$ be the vector with all entries equal $1$. Solve the linear program: $\max \sum\limits_{i=1}^{n} x_i-1$, s.t. $Ax\leq 1_d, x\geq 0$. The reciprocal of the optimal value is the cut-set bound.
\end{itemize}
\end{algorithm}

When routing is performed, information is transmitted from S to D without being modified or coded at R.  {However, to protect the message, one needs to perform coding at S and decoding at D. Otherwise, if  $M$ is not coded along with $K$, then
$M$ cannot be secure.} Hence, we may assume there are $n_1\times n_2$ paths/channels directly connecting S to D.
Since information is unchanged in each path, a wiretapper who can access channel $e$ will know the information on all the paths that pass through $e$. Hence we need to define a new point-to-point communication system (or cut-set) and the corresponding wiretap patterns.
The cut-set $W'=\{e_{1,1}',e_{1,2}',...,e_{n_1,n_2}'\}$. The corresponding wiretap pattern $\mathcal{A}'$ is constructed as follows: For each $A\in \mathcal{A}$, there is a  corresponding $A'\in \mathcal{A}'$ such that $A' = \{e_{i,k}' (1\leq k\leq n_2): e_{1,i}\in A\}\cup \{e_{k,j}' (1\leq k\leq n_1): e_{2,j}\in A\}$.
Then the routing bound can be computed by applying the cut-set bound on $W'$ and $\mathcal{A}'$. {According to \cite{ChengYeung2011NetC, CY2014IT}, if the cut-set bound is achieved, one can also recover $K$ from the specified coding scheme. Thus routing bounds are also the same for both  Class-I and Class-II $(n_1,n_2)$ networks.} {The algorithm is described in Alg. 2.}
\begin{algorithm}\label{algorithm:2}
\caption{Algorithm for computing the routing bound}
\begin{itemize}
 \item [1] Denote the number of wiretap sets by $d$.  For ease of discussion, use $1,2,..., n_1$ and $1, 2, ...,n_2$ to index the edges in the two layers of the network.
 \item [2] There are $n = n_1\times n_2$ paths in the routing scheme. All the paths are ordered as follows.  {If a path  goes through the $i$th edge in layer $1$ and $j$th edge in layer $2$ of the network, then its index is $(i-1)\times n_2 + j$.} Let $A$ be a $d \times n$ matrix.
   \item [3] If the $i$th wiretapper can access the $j$th path, then $A(i, j) = 1$. Otherwise, $A(i,j) = 0$.
 \item [4] Let $1_d$ be a vector with all entries equal $1$. Solve the linear program: $\max \sum\limits_{i=1}^{n} x_i-1$, s.t. $Ax\leq 1_d, x\geq 0$. The reciprocal of the optimal value is the routing bound.
\end{itemize}
\end{algorithm}

The idea of Shannon bound is from Yeung \cite[Ch. 14]{yeung2008information}, where the key principle is to transform the problem  to a linear program in the two elemental forms; i.e.,
\begin{itemize}
\item [(i)]$ H(X_i|X_{[n]-\{i\}}), i\in [n]$;
\item [(ii)] $I(X_i;X_j|X_{K})$, where $i\neq j$ and $K\subseteq [n]-\{i,j\}$.
\end{itemize}
The  algorithm for computing Shannon bound has been already implemented in ITIP/Xitip. {When we input all the constraints, ITIP/Xitip will tell us whether $H(K)\geq c H(M)$ is true or false. After gradually adjusting $c$, the Shannon bound will be obtained.}

Intuitively, we may speculate that routing is optimal for many wiretap patterns. But it is not trivial to prove its optimality. For Class-I/II $(2$, $2)$ networks, we can check that routing is optimal. {For general $(n_1,n_2)$ networks, numerical experimentations  by  computer programs are preferred.
For each wiretap pattern $\mathcal{A}$, denote
\begin{equation}
\tau_{\mathcal{A}}=\min\frac{H(K)}{H(M)},
\end{equation}
where the minimum is taken over all coding schemes (per Sec. \ref{sec1}).   Recall that  various bounds in the literature satisfy that:
\begin{equation}\label{bound-chain}
\text{Cut-set bound }  \leq \text{Shannon bound } \leq \tau_{\mathcal{A}} \leq \text{ Linear Network Coding bound} \leq \text{ Routing bound.}
\end{equation}
{The linear network coding bound above is the bound when optimized over all linear network coding schemes.}
In principle, if any lower bounds match  any upper bounds, then  $\tau_{\mathcal{A}}$ is determined.}  Otherwise,  further investigations must be conducted on such wiretap patterns. In the following, we conduct computation-based experiments to study the optimality of routing in Class-I/II networks.

\section{On the sub-optimality of routing}

\subsection{Experiment}\label{sec2}
When using ITIP/Xitip to assess the optimality of routing on the Class-I/II $(n_1,n_2)$ networks, we face the challenge of having to deal with an immense computational complexity. In Class-I/II $(n_1,n_2)$ networks, since the wiretap pattern $\mathcal{A}$ is a subset of $2^{\mathcal{E}}$, in essence, we need to exhaust the set of all possible wiretap patterns, whose size is $2^{2^{|\mathcal{E}|}}= 2^{2^{n_1+n_2}}$. Therefore, in principle, when $n_1+n_2>6$, it is hard to enumerate all patterns. The other difficulty is from the computational complexity of ITIP/Xitip, because we need to solve a linear program with $\binom{n_1+n_2+2}{2}2^{n_1+n_2}+(n_1+n_2+2)$ variables.

For a given wiretap pattern $\mathcal{A}$, we may assume that  no element  is a subset of another element. In this manner, the number of different wiretap patterns for the Class-I/II $(n_1,n_2)$ networks can be  largely reduced.
For a given set $P$, a set $V\subseteq 2^{P}$ is called an \textit{antichain} if it satisfies that for all different $v_1\in V$ and $v_2\in V$,  $v_1  \not\subseteq v_2$ and  $v_2 \not\subseteq v_1$. The number of antichains for a set with size $n=0,1,2,...,$ is listed in Table \ref{tab-anti} (see Sloane \cite{no-anti-chain}).
\begin{table}[!htb]
\begin{center}
\begin{tabular}{|c|c|c|c|c|c|c|c|c|c|}
      \hline
      |$n$|  & 0 & 1 & 2 & 3  & 4   & 5 & 6 & 7  \\ \hline
      \#  & 2 & 3 & 6 & 20 & 168 & 7,581 & $7,828,354$ & $2,414,682,040,998$ \\
      \hline
    \end{tabular}
    \caption{The number of antichains}
    \label{tab-anti}
\end{center}
\end{table}
In light of the  computational complexity, $n=n_1+n_2 = 6$ is the limiting problem size for ITIP/Xitip based algorithm. In the following,  we only focus on the symmetric case where $n_1= n_2=3$. 

After reducing  the problem size, we design the algorithm in Alg. 3:
\begin{algorithm}\label{algorithm:simu}
\caption{Algorithm for assessing the tightness of the routing bound}
Generate all the possible wiretap patterns which are antichains. For each wiretap pattern,
\begin{itemize}
\item [1.] Compute the cut-set bounds on $(S,R)$ and $(R,D)$, respectively. Keep the larger one and denote it by $l_1$.
\item [2.] Compute the routing bound on $S\to R\to D$. Denote it by $l_2$.
\item [3.] If $l_1$ equals $l_2$, then the routing bound is tight. And hence Shannon bound is tight too. Proceed to Step $5$.
\item [4.] Compare $l_2$ with the Shannon bound in ITIP/Xitip. If these two bounds are equal, then routing is optimal; otherwise, there is a gap between routing bound and Shannon bound.
\item [5.] Proceed to the next wiretap pattern.
\end{itemize}
\end{algorithm}

Steps $1$ and $2$ are used to reduce computational cost, since computing the cut-set bound is much faster than the Shannon bound.
The experiment lasted for around 3--4 days.  We keep track of some records during the experiment. For almost 80$\%$ of the wiretap patterns, the cut-set bounds match  the routing bounds. In the Class-I $(3,3)$ network, there are around $159,258$ wiretap patterns ($2\%$ of all the  wiretap patterns) where gaps between routing bounds and Shannon bounds are found.  In the Class-II $(3,3)$ network, there are around $32,472$ wiretap patterns (0.4$\%$ of all the  wiretap patterns) where gaps exist.

When a gap is found between the  routing bound and the Shannon bound, there are two cases: one is that the routing bound is not tight, and network coding is needed; the other is that Shannon-type inequalities are not sufficient, and non-Shannon-type inequalities should be used. Next,  some wiretap patterns are  analyzed to show that routing is not sufficient for the Class-I/II (3, 3) network, and network coding is necessary to achieve $\min H(K)/H(M)$.

\subsection{Simplification of search procedure}\label{sec3}

In the following sections, we choose some wiretap patterns to demonstrate the advantage of coding over routing, and the distinction between the Class-I and Class-II networks. Since the problem size is small, we just use $X_1, \ldots ,  X_6$ and $e_1\ldots,e_6$ to denote the random variables and edges. {For ease of verification, we provide with the code written in \Xitip\ and Matlab in~\cite{cvcode1}. } 

From existing works of the literature, we do not know whether Shannon bound is tight or linear network coding is sufficient.  {However, since the network topology is  simple,  we may first assume that the Shannon bound is tight and linear network codes suffice to achieve the Shannon bound. As such, we attempt to construct such a linear network code.} To the best of our knowledge, there is no systematic and well-established technique  on how to achieve the Shannon bound. One possible method is to study the structure of the optimal solution resulting from the linear program of the Shannon bound, by which some properties about the coding scheme may be found. This approach suffers from the curse of high dimensionality of the optimal solution. Here we provide an alternative method. Note that if we add more constraints to a linear program of the form
\begin{equation}
\max c^Tx,\quad \mbox{s.t.}\quad Ax\leq b,
\end{equation}
 the optimal value will not be increased. If the optimal value remains the same, then we can use these additional  constraints to  reduce the search space for an optimal solution. If we use this technique to study the Shannon bound, some  properties about the coding scheme may be obtained.

The first  simple intuition is that if we add the following constrains into the linear programs, the Shannon bounds always remain the same:
\begin{align}
H(X_1, X_2, X_3) = H(X_1) + H(X_2) + H(X_3)\\
H(X_4, X_5, X_6) = H(X_4) + H(X_5) + H(X_6)
\end{align}
That is the random variables on the same layer are constrained to be mutually independent.

Another idea is based on the observation that for a coding scheme, the following holds:
\begin{equation}
H(X_4,X_5, X_6|X_1, X_2, X_3) = 0. \label{funcRelation}
\end{equation}
 This relationship is rather general since the functional relationships between the random variables may be more precise; e.g., $X_4$ may be a function of $X_1$ and $X_2$, and $X_3$ is not involved in the coding. The exact functional relationships will be very helpful in the construction of a linear coding scheme. Our algorithm is as follows: For each $X_i$ ($4\leq i\leq 6$), enumerate all the possible functional relationships with the subsets of $X_1, X_2, X_3$. Then update the functional relationship and check by ITIP/\Xitip\ to see whether the Shannon bound is unchanged. If so, then the functional relationship is valid and does not decrease the optimal value of the linear program. Here we state an example to illustrate our approach.

{Before  doing so, let us comment about the alphabet of the underlying random variables. As we mentioned in the definition of the code (in Section \ref{sec1}), we are allowed to choose the alphabets. To keep the design of the codes simple, the alphabets henceforth are kept fixed  to be $\mathbb{F}_q$ (for some $q\geq 3$) in the coding schemes we design. In the sequel, unless otherwise stated, all the alphabets  are  the finite field $\mathbb{F}_q$.}

\begin{Example}
In the Class-II $(3,3)$ network (depicted in Fig. \ref{general-model-33}), let $\mathcal{A} = \{A_1, A_2, A_3, A_4, A_5\}$, where $A_1=\{e_2, e_4, e_5\}$, $A_2=\{e_2, e_3, e_6\}$, $A_3=\{e_1, e_5, e_6\}$, $A_4=\{e_1, e_3, e_4\}$, and $A_5=\{e_1, e_2, e_4, e_6\}$. The routing bound is equal to $4$. By the following constraints, the Shannon bound is  equal to $3$.
\begin{align} 
I(M;K) = 0 \label{ex1:con1}\\
H(X_1, X_2, X_3| M ,K ) = 0 \\
H(X_4, X_5, X_6|X_1, X_2, X_3) =0\\
H(M,K|X_4, X_5, X_6) = 0\\
I(M; X_2, X_4, X_5) = 0\\
I(M; X_2, X_3, X_6) = 0\\
I(M; X_1, X_5, X_6) = 0\\
I(M; X_1, X_3, X_4) = 0\\
I(M; X_1, X_2, X_4,X_6) = 0 \label{ex1:con2}
\end{align}
After adding the  following constraints, the Shannon bound remains the same.
\begin{align}
H(X_1, X_2, X_3) = H(X_1) + H(X_2) + H(X_3) \label{ex1:con3}\\
H(X_4, X_5, X_6) = H(X_4) + H(X_5) + H(X_6)
\end{align}
After checking with ITIP/Xitip, we obtain these additional functional relationships:
\begin{align}
H(X_4|X_1,X_3)  =0\\
H(X_5| X_1, X_2,X_3) =0\\
H(X_6|X_2,X_3) = 0\label{ex1:con4}
\end{align}
The Shannon bound will be changed if we add
\begin{align}
H(X_1) = H(X_2) = H(X_3) \\
H(X_4) = H(X_5) = H(X_6)
\end{align}
After checking with ITIP/Xitip, we find that the following constraints can be added  without changing the Shannon bound.
\begin{align}
& H(X_3) = 2H(X_1) \label{ex1:con5}\\
& H(X_1) = H(X_2)\\
& H(X_5) = 2H(X_4)\\
& H(X_4) = H(X_6) \label{ex1:con6}
\end{align}
From  these constraints, we see that the information rates on the edges from S to R  and R to D are not identical any more.

The coding scheme is constructed as follows: Split $X_3$ into $(X_{31}, X_{32})$ and $X_5$ into $(X_{51}, X_{52})$.
\begin{itemize}
\item [1.] Independently  generate three bits of key $K_1$,  $K_2$ and $K_3$ and one bit of message $M$ from  $\mathbb{F}_q$;
\item [2.] on edge $e_1$, $X_1 = M + K_1 + K_2 + K_3$;
\item [3.] on edge $e_2$, $X_2 = K_1$;
\item [4.] on edge $e_3$, $X_{31}= K_2$ and $X_{32}= K_3$;
\item [5.] at intermediate node R, $(M, K_1,K_2,K_3)$ can be easily recovered;
\item [6.] on edge $e_4$, $X_4 =  X_1+X_{31} =  M + K_1 + 2K_2 + K_3$;
\item [7.] on edge $e_5$, $X_{51}=  X_1 +X_2 +X_{31} = M + 2K_1 + 2K_2 + K_3$ and $X_{52} = X_1 + X_2 +X_{32} = M + 2K_1 + K_2 + 2K_3$;
\item [8.] on edge $e_6$, $X_{6} = X_2 + X_{31} = K_1 + K_2$;
\item [9.] at the destination node D, $(X_1, X_2, X_{31},X_{32})$ can be recovered by $X_1 =  X_{51}-X_6$,  $X_2 = X_{51}-X_4$, $X_{31} = X_6 - X_{51}+X_4$, and  $X_{32} = X_{52} +X_4+X_6- 2X_{51}  $,  which means $(M, K_1, K_2, K_3)$ can also be recovered.
\end{itemize}
We now verify the security constraints.
\begin{itemize}
\item $A_1$: $I(M; X_2, X_4,X_5) \\
= I(M; X_2, X_1+X_{31}, X_1+X_2+X_{31}, X_1+X_2+X_{32} )\\
= I(M; X_2, X_1+X_{31}, X_1+X_{32}) \\
= H(X_2, X_1+X_{31}, X_1+X_{32}) - H(X_2, X_1+X_{31}, X_1+X_{32}|M)\\
= H(X_2, X_1+X_{31}, X_1+X_{32}) - H(K_1, K_1 +2K_2+K_3, K_1 + K_2 +2K_3)\\
= 0$
\item $A_2$: $I(M; X_2, X_3, X_6)\\
= I(M; X_2, X_3, X_2 +X_{31})  \\
= I(M; X_2, X_3) \\
= 0$
\item $A_3$: $I(M; X_1, X_5, X_6) \\
= I(M;X_1,  X_1+X_2 +X_{31}, X_1+X_2 +X_{32}, X_2 +X_{31} )\\
= I(M;X_1,  X_2 +X_{31}, X_2 +X_{32}  )  \\
= 0$
\item $A_4$: $I(M; X_1, X_3, X_4) \\
= I(M; X_1, X_3, X_1 +X_{31}) \\
= I(M; X_1, X_3)\\
= 0$
\item $A_5$: $I(M; X_1, X_2, X_4, X_6) \\
= I(M; X_1, X_2, X_1+X_{31}, X_2 +X_{31})\\
=I(M; X_1, X_2, X_{31})\\
=0$
\end{itemize}
 An   interesting discovery is that, in ITIP/Xitip,   if we set $H(K) = 3H(M)$ together with (\ref{ex1:con1})-(\ref{ex1:con2}) in the constraints, all the relationships in (\ref{ex1:con3})-(\ref{ex1:con4})  and (\ref{ex1:con5})-(\ref{ex1:con6}) are true. This means that if the ratio of the size of the key and the message is fixed to that prescribed by the Shannon bound, then several functional relationships between the constituent random variables in the problems can be uncovered. {In fact,  these functional relationships can help the code designer to find an optimal linear code. For example \eqref{ex1:con3} tells us that $X_1, X_2$ and $X_3$ must be independent in any optimal coding scheme. } To the best of the authors' knowledge, this observation has not been   made in previous studies.    
\end{Example}
{ An algorithm based on the heuristic observation above is described in Alg. \ref{algorithm:20151111}.
\begin{algorithm}
\caption{Algorithm for finding the hidden structure of coding schemes}\label{algorithm:20151111}
In ITIP/Xitip, input all the constraints from the problem setting (e.g., \eqref{ex1:con1}-\eqref{ex1:con2}).
\begin{itemize}
\item [1.] Add the Shannon bound $H(K) = c H(M)$ in the constraints in ITIP/Xitip, where $c$ is the Shannon bound.
\item [2.] Check whether $H(X_1, X_2, X_3) = H(X_1) + H( X_2) + H( X_3)$ and  $H(X_4, X_5, X_6) = H(X_4) + H( X_5) + H( X_6)$ are true.
\item [3.] Check the ratios of $ H(X_1) : H( X_2) : H( X_3) $ and $H(X_4):H( X_5): H( X_6)$. One may first find  that $H(X_1)=c_1 H(X_2)$ and $H(X_1)=c_2 H(X_3)$. To find $c_1$ such that $H(X_1)=c_1 H(X_2)$, it is equivalent to find $c_1$ such that $H(X_1)\geq c_1 H(X_2)$ and $H(X_1)\leq c_1 H(X_2)$, which can be checked by ITIP/Xitip.
\item [4.] Check the functional relationships between $(X_1, X_2, X_3)\to (X_4, X_5, X_6)$ (e.g., $X_4$ is a function of ($X_1$, $X_2$) denoted as  $(X_1,X_2)\to X_4$).
\end{itemize}
\end{algorithm}}

{All the additional conditions can help us to construct an optimal linear coding scheme.}

\begin{Remark}\label{rmk:1}
{In principle, to design a linear network code, it suffices to associate a vector space to each channel in the $(3,3)$ network. These six vector spaces can be discussed in the context of the linear entropic region of 6 vector spaces, namely,  $\tilde{L}_6$. However, the theory of determining the structure of $\tilde{L}_6$ is incomplete. This hampers its application to our problem. }
\end{Remark}
\subsection{Coding is necessary} \label{cin}

\begin{Example}
In the Class-I/II $(3,3)$ network (depicted in Fig. \ref{general-model-33}), consider the wiretap pattern $\mathcal{A}$ $=$ $\{A_1$, $A_2$, $A_3$, $A_4\}$,
where $A_1=\{2,3,5\}$, $A_2=\{1,4,5\}$, $A_3=\{1,3,6\}$, and $A_4=\{2,4,6\}$. The routing bound is equal to $3$. Both of the Shannon bounds for the Class-I and Class-II settings are equal to $2$. Next, we construct a code to demonstrate that $2$ is optimal.

\begin{itemize}
\item [1.] Independently  generate two bits of key $K_1$ and $K_2$ and one bit of message $M$ from  $\mathbb{F}_q$;
\item [2.] on edge $e_1$, $X_1 = K_1$;
\item [3.] on edge $e_2$, $X_2 = K_2$;
\item [4.] on edge $e_3$, $X_3= M+K_1+K_2$;
\item [5.] at intermediate node R, $(M, K_1,K_2)$ can be easily recovered;
\item [6.] on edge $e_4$, $X_4 = M +2K_1+2K_2$;
\item [7.] on edge $e_5$, $X_5 = M + K_1+2K_2$;
\item [8.] on edge $e_6$, $X_6 = M + 2K_1 + K_2$;
\item [9.] at the destination node D, $(M, K_1, K_2)$ can be recovered by $M =  2(X_5+X_6)-3X_4$, $K_1 = X_4-X_5$, and $K_2 = X_4-X_6$.
\end{itemize}
We now verify the security constraints.
\begin{itemize}
\item $A_1$: $I(X_2,X_3,X_5;M)\\
 = H(X_2,X_3,X_5) - H(X_2,X_3,X_5|M) \\
=  H(K_2, M+K_1+K_2, M+K_1+2K_2)\\
   - H(K_2,M+K_1+K_2,M+K_1+2K_2|M) \\
= H(K_2,M+K_1) - H(K_2,K_1|M) \\
= 0$
\item $A_2$: $I(X_1,X_4,X_5;M) \\
= H(X_1,X_4,X_5) - H(X_1,X_4,X_5|M) \\
=  H(K_1, M+2K_1+2K_2, M+K_1+2K_2) \\
- H(K_1, M+2K_1+2K_2, M+K_1+2K_2|M) \\
= H(K_1, M+2K_2) - H(K_1, K_2|M) \\
= 0$
\item $A_3$: $I(X_1,X_3,X_6;M) \\
= H(X_1,X_3,X_6) - H(X_1,X_3,X_6|M) \\
=  H(K_1, M+K_1+K_2, M+2K_1+K_2)\\
- H(K_1, M+K_1+K_2, M+2K_1+K_2|M) \\
= H(K_1,M+K_2) - H(K_1,K_2|M) \\
= 0$
\item $A_4$: $I(X_2,X_4,X_6;M) \\
= H(X_2,X_4,X_6) - H(X_2,X_4,X_6|M)\\
=  H(K_2, M+2K_1+2K_2, M+2K_1+K_2) \\
- H(K_2, M+2K_1+2K_2, M+2K_1+K_2|M) \\
= H(K_2, M+2K_1) - H(K_2, K_1|M) \\
= 0$
\end{itemize}
\end{Example}
\begin{Remark}
To design a secure linear coding scheme, one needs to design $X_i$, $i =1$, ..., $6$, such that $X_1$, $X_2$, and $X_3$ are mutually independent and  $X_4$, $X_5$, and $X_6$ are also mutually independent. Since we have two bits for the key and one bit for the message, the information rates in a wiretap set cannot exceed  two bits. On one hand, to reduce the  size of key, observe from the wiretap set $A_1$ that we need to ensure that $X_5$ is a  function of $X_2$ and $X_3$. On the other hand, since $(X_1, X_2, X_3)$ is recoverable from $(X_4, X_5, X_6)$, the information rates of $X_4$, $X_5$, $X_6$ should be $3$. There  should be a trade-off between these two constraints, which may make routing sub-optimal.
\end{Remark}

 \subsection{Distinction between Class-I and Class-II Networks}
 \begin{Example}
In this example, we  show that for the same network topology and wiretap pattern, the routing bound is tight for the  Class-II network while it is loose for the Class-I network.
Consider the wiretap pattern $\mathcal{A} = \{A_1, A_2, A_3, A_4\}$, where $A_1=\{1,4\}$, $A_2=\{2,3,4\}$, $A_3=\{1,2,5,6\}$, and $A_4=\{3,5,6\}$. The routing bound is $3$. The Shannon bound in Class-I network is $2$ and in Class-II network is $3$. Next, we  construct a code to show that $2$ is optimal for the Class-I network.
\begin{itemize}
\item [1.] Independently generate two bits of key $K_1$ and $K_2$ and one bit of message $M$ from $\mathbb{F}_q$;
\item [2.] on edge $e_1$, $X_1 = M+K_1$;
\item [3.] on edge $e_2$, $X_2 = K_2$;
\item [4.] on edge $e_3$, $X_3= K_1$;
\item [5.] it is easy to see at intermediate node R, $(M, K_1, K_2)$ can be recovered.
\item [6.] on edge $e_4$, $X_4 = K_1+K_2$;
\item [7.] on edge $e_5$, $X_5 = M+K_1+K_2$;
\item [8.] on edge $e_6$, transmit nothing;
\item [9.] at the destination node D, $M$ can be recovered by $M =  X_5-X_4$, $K_1$ and $K_2$ cannot be recovered.
\end{itemize}
We now verify the security constraints.
\begin{itemize}
\item $A_1$: $I(X_1,X_4;M) \\
= H(X_1,X_4) - H(X_1,X_4|M) \\
=  H(M+K_1, K_1+K_2) - H(M+K_1, K_1+K_2|M) \\
= H(M+K_1, K_1+K_2) - H(K_1,K_2|M) \\
= 0$
\item $A_2$: $I(X_2,X_3,X_4;M) \\
= I(K_2,K_1, K_1+K_2;M)\\
= 0$
\item $A_3$: $I(X_1,X_2,X_5,X_6;M) \\
= H(X_1,X_2,X_5,X_6) - H(X_1,X_2,X_5,X_6|M) \\
=  H(M+K_1, K_2, M+K_1+K_2) \\
- H(M+K_1, K_2, M+K_1+K_2|M) \\
= H(M+K_1,K_2) - H(K_1, K_2|M) \\
= 0$
\item $A_4$: $I(X_3,X_5,X_6;M) \\
= H(X_3,X_5,X_6) - H(X_3,X_5,X_6|M) \\
=  H(K_1, M+K_1+K_2) - H(K_1, M+K_1+K_2|M) \\
= H(K_1, M+K_2) - H(K_1, K_2|M) \\
= 0$
\end{itemize}
\end{Example}

\subsection{Sub-optimality of routing in Class-I/II $(n_1,2)$ and $(2,n_2)$ networks}\label{subop32}

 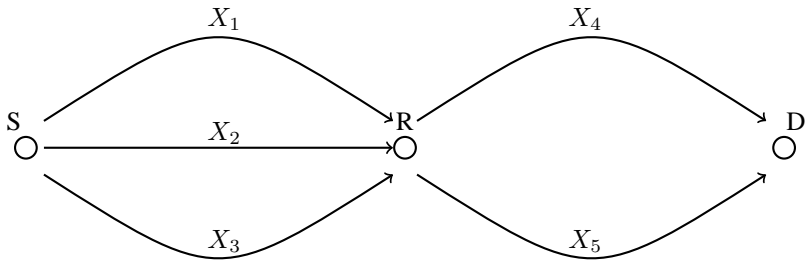
\begin{figure}
  \centering
\begin{tikzpicture}[xscale=0.8,yscale=0.8]

\draw  [ thick](-0.3,1.854) circle [radius=0.18];
\draw  [ thick](6,1.854) circle [radius=0.18];
\draw  [ thick](12+0.3,1.854) circle [radius=0.18];

\draw [thick, ->] (0,1.854*2-2.3) .. controls (2.9 ,1.854-2.3)  .. (6-0.2, 1.854*2-2.3);
\draw [thick, ->] (6+0.2, 1.854*2-2.3) .. controls (9.1 ,1.854-2.3)  .. (12, 1.854*2-2.3);

\draw [ thick, ->] (0,1.854) -- (6-0.2,1.854);

\draw [thick, ->] (0,2.3) .. controls (2.9 ,1.854+2.3)  .. (6-0.2, 2.3);
\draw [thick, ->] (6+0.2, 2.3) .. controls (9.1 ,1.854+2.3)  .. (12, 2.3);

\node at (-0.5, 2.3) {S};
\node at (6, 2.3) {R};
\node at (12+0.5, 2.3) {D};

\node at (1.5*2, 4) {$X_{1}$};
\node at (1.5*2, 2.1) {$X_{2}$};
\node at (1.5*2, 0.3) {$X_{3}$};
\node at (4.5*2, 4) {$X_{4}$};
\node at (4.5*2, 0.3) {$X_{5}$};
%
%
%
%
%
\end{tikzpicture}

  \caption{Class-I/II (3, 2) network}\label{3-2-model}
\end{figure}
\begin{Example}
The Class-II ($n_1$, $n_2$) network is equivalent to the Class-II ($n_2$, $n_1$) network. For Class-I networks, the situation is more subtle. For Class-I/II $(2,n_2)$ networks,  experiments can  show that routing is optimal when $n_2\leq 3$. The problem is  still open for the case $n_2 \geq 4$. In the following, we show that there exists a Class-I $(3,2)$ network (depicted in Fig. \ref{3-2-model}), where routing is not optimal.
 Consider the following wiretap pattern $\mathcal{A} = \{A_1,A_2,A_3,A_4\}$, where $A_1=\{1,2,4\}$, $A_2=\{3,4\}$, $A_3=\{2,5\}$, and $A_4=\{1,3,5\}$. The routing bound is  equal to $3$. The Shannon bound for Class-I is equal to $2$ and for Class-II is equal to $3$. Next, we  construct a code to show that $2$ is optimal for the Class-I network.
\begin{itemize}
\item [1.] Independently generate two bits of key $K_1$ and $K_2$ and one bit of message $M$ from $\mathbb{F}_q$;
\item [2.] on edge $e_1$, $X_1 = K_1$;
\item [3.] on edge $e_2$, $X_2 = M+K_2$;
\item [4.] on edge $e_3$, $X_3= K_2$;
\item [5.] it is easy to see at intermediate node R, $(M,K_1,K_2)$ can be recovered.
\item [6.] on edge $e_4$, $X_4 = M+K_1+K_2$;
\item [7.] on edge $e_5$, $X_5 = K_1+K_2$;
\item [8.] at the destination node D, $M$ can be recovered by $M =  X_4-X_5$, $K_1$ and $K_2$ cannot be recovered.
\end{itemize}
{The security constraints can be readily verified by the same approach in   Examples $1$ to $3$.}
\end{Example}




\section{Two Hard Examples}\label{hdexam}
In the previous sections, we have already elucidated the differences between Class-I and Class-II wiretap networks. In the sequel, we focus on the Class-II wiretap networks since the problem may become simpler when $K$ is required to be recovered.  An observation from the experiments in the previous section is that   gaps  between a lower bound and an upper bound in \eqref{bound-chain} exist only if $4 \leq |\mathcal{A}| \leq 12$. {For $|\mathcal{A}|= 4,\ldots, 12$, the number of wiretap patterns where gaps exist is listed in Table II.}

\begin{table}[!htb]
\begin{center}

\begin{tabular}{|c|c|c|c|c|c|c|c|c|c|}
  \hline
  $|\mathcal{A}|$ & 4    & 5    & 6    & 7    & 8   & 9      & 10 & 11 & 12 \\ \hline
  \text{\# gaps}              & 18   & 252 & 1494 & 4842 & 9144 & 9648 & 5400 & 1494 & 180\\
  \hline
\end{tabular}
\caption{the number of wiretap patterns where gaps exist in the Class-II network}
\end{center}
\end{table}

For wiretap patterns in which the routing bounds are not  optimal, we have constructed linear coding schemes to achieve the Shannon bounds by hand. This may be tedious and non-systematic.  It is thus of great interest to see whether we can apply similar techniques to all the remaining unknown cases. {When $|\mathcal{A}|=4$, there are only $18$ wiretap patterns, where all the routing bounds are equal to $3$ and Shannon bounds are equal to $2$.} {We can check that  Shannon bounds  for all these cases can be achieved by linear  coding schemes by the method {(Alg. \ref{algorithm:20151111})} in Sec.\ \ref{sec3}. We present a similar  coding scheme in Sec. \ref{cin}.}  For wiretap patterns where $|\mathcal{A}|>4$ there are instances for which optimal linear coding schemes are not easy to construct by hand. Next, we  state some wiretap patterns to demonstrate the difficulty of these specific instances. 

\begin{Example}
Let $\mathcal{A} = \{A_1, A_2, A_3, A_4, A_5\}$, where $A_1=\{e_2, e_4, e_5\}$, $A_2=\{e_2, e_3, e_6\}$, $A_3=\{e_1, e_4, e_6\}$, $A_4=\{e_1, e_3, e_5\}$, and $A_5=\{e_1, e_2, e_4\}$. The routing bound is equal to 3. By the following constraints, the Shannon bound is  equal to $7/3$.
\begin{align}
3H(K) \geq 7 H(M) \\
I(M;K) = 0 \\
H(X_1, X_2, X_3| M ,K ) = 0 \\
H(X_4, X_5, X_6|X_1, X_2, X_3) =0 \\
H(M,K|X_4, X_5, X_6) = 0 \\
I(M; X_2, X_4, X_5) = 0 \\
I(M; X_2, X_3, X_6) = 0 \\
I(M; X_1, X_4, X_6) = 0 \\
I(M; X_1, X_3, X_5) = 0 \\
I(M; X_1, X_2, X_4) = 0
\end{align}
The following constraints  can be added without changing the Shannon bound.
\begin{align}
H(X_1,X_2,X_3) = H(X_1) + H(X_2) + H(X_3) \\
H(X_4,X_5,X_6) = H(X_4) + H(X_5) + H(X_6) \\
H(X_1) = H(X_2) \\
3H(X_3) = 4H(X_1)\\
H( X_5 ) = H( X_6 ) \\
2H( X_4 ) = H( X_5 ) \\
H( X_4 | X_1, X_2, X_3 ) = 0\\
H( X_5 | X_1, X_3 ) = 0 \\
H( X_6 | X_2, X_3 ) = 0
\end{align}

Since $H(K)/H(M) = 7/3$, we need to construct a linear code on a vector with size $10$ and the bit rates on each edges from S to R are $3$, $3$, and $4$ and from R to D are $2$, $4$, and $4$, respectively. That is tantamount to  a $10 \times 10$ matrix over $\mathbb{F}_q$. Except brute force search, we have no other choice at the moment. Since the space of the feasible  solutions is  huge,  optimal linear coding schemes for this example are unknown.
\end{Example}

Next,  an even more complicated wiretap pattern with $12$ wiretap sets is studied.
\begin{Example}
Let $\mathcal{A} = \{A_1$, $A_2$, ..., $ A_{12}\}$,
where $A_1=\{e_3, e_5, e_6\}$, $A_2=\{e_3, e_4, e_6\}$, $A_3=\{e_3, e_4, e_5\}$, $A_4=\{e_2, e_5, e_6\}$, $A_5=\{e_2, e_4, e_6\}$, $A_6=\{e_2, e_3, e_6\}$, $A_7=\{e_2, e_3, e_5\}$, $A_8=\{e_2, e_3, e_4\}$, $A_9=\{e_1, e_5, e_6\}$, $A_{10}=\{e_1, e_3, e_5\}$, $A_{11}=\{e_1, e_3, e_4\}$, and $A_{12}=\{e_1, e_2, e_4, e_5\}$. The routing bound is equal to 4. The Shannon bound is equal to 19/5 by the following constraints.

\begin{align}
5H(K) \ge 19 H(M)\\
I(M;K) = 0\\
H(X_1, X_2, X_3| M ,K ) = 0\\
H(X_4, X_5, X_6|X_1, X_2, X_3) =0\\
H(M,K|X_4, X_5, X_6) = 0 \\
I(M; X_3, X_5, X_6) = 0\\
I(M; X_3, X_4, X_6) = 0\\
I(M; X_3, X_4, X_5) = 0\\
I(M; X_2, X_5, X_6) = 0\\
I(M; X_2, X_4, X_6) = 0\\
I(M; X_2, X_3, X_6) = 0\\
I(M; X_2, X_3, X_5) = 0\\
I(M; X_2, X_3, X_4) = 0\\
I(M; X_1, X_5, X_6) = 0\\
I(M; X_1, X_3, X_5) = 0\\
I(M; X_1, X_3, X_4) = 0\\
I(M; X_1, X_2, X_4, X_5) = 0
\end{align}
We can show via ITIP/Xitip and hence prove that the Shannon bound is  unchanged after adding the following constraints.
\begin{align}
H(X_1, X_2, X_3) = H(X_1) + H(X_2) + H(X_3)\\
H(X_4, X_5, X_6) = H(X_4) + H(X_5) + H(X_6)\\
7H(X_1) = 9H(X_2)\\
8H(X_1) = 9H(X_3)\\
 3 H( X_4 ) = 4 H( X_5 ) \\
        5 H( X_4 ) = 4 H( X_6 ) \\
	H( X_4 | X_1, X_2, X_3 ) = 0 \\
	H( X_5 | X_1, X_2, X_3 ) = 0 \\
	H( X_6 | X_1, X_2, X_3 ) = 0
\end{align}
Therefore, we need  to define  a linear  coding scheme on a vector with size $24$ such that $H(K) = 19$ and $H(M) = 5$. Moreover, the bit rates on the layers from S to R are $9$, $7$, and $8$, and from R to D are 8, 6, and 10,  respectively. The functional relationships between random variables cannot be further refined. Hence, just as in Example 1, we need to split each of the random variables $X_1$, $X_2$, ..., $X_6$ (say $X_1$ into ($X_{11}$, $X_{12}$, ..., $X_{19}$)), then by ITIP/Xitip, we can obtain more refined  relationships between  constituent random variables $\{X_{ij}\}$. Due to the number of random variables (i.e., 24), ITIP/Xitip cannot afford such a huge computation. The linear coding schemes are unknown.
\end{Example}

{In principle, to find $\tau_{\mathcal{A}}$, one needs  to find  upper and lower bounds that match.
 Since the network model is simple, it is plausible that the Shannon bound is tight (i.e., it is equal to $\tau_{\mathcal{A}}$) and linear network coding  schemes are sufficient.} From these two examples, we observe that we do not have efficient methods to construct linear coding schemes to achieve their respective Shannon bounds in general. The only method as of now is to use a brute force search over all the possible linear network coding schemes over $\mathbb{F}_q$. The computational complexity of such a search is  $q^{n^2}$, where $n$ is the size of the vector and $q$ is the field size. In light of the number of  the remaining wiretap patterns,    it is intractable and computationally prohibitive  if we simply rely on the state-of-the-art resources. Even if we can find a solution by brute force search, such a solution fails to provide us with an intuitive understanding of the structure of the coding scheme. 
Thus, one of the take-home messages in this paper is that information theorists need to construct new techniques and concepts to systematically achieve the Shannon bound using linear coding schemes. Without such techniques, it is difficult to make progress. {To assist the readers of this paper to investigate the problem further, we have provided the data in \cite{cvcode2}.}

\section{Conclusion}\label{sec4}
In this paper, we have defined Class-I/II ($n_1$, $n_2$) wiretap networks and have numerically computed bounds on $H(K)/H(M)$. The performances of the routing bound on various wiretap patterns have been compared to the cut-set bound and the Shannon bound.  Examples are constructed to show that under both decoding classes, routing is not sufficient even for a simple (3, 3) network. Our numerical calculations also demonstrate the differences between   Class-I    and Class-II networks.  { Our study shows that when $|\mathcal{A}|<4$ or $|\mathcal{A}|>12$, routing is optimal. Gaps exist in other cases. In the ($3$, $3$) Class-II wiretap network, we can construct a linear code to achieve the Shannon bound for some wiretap patterns. When $5\leq |\mathcal{A}|\leq 12$, it is an open problem to systematically construct optimal linear codes, though we have provided some heuristic rules. A summary on the examples studied is provided in Table \ref{tab:3}.}
\begin{table}[!htb]
\begin{center}
\begin{tabular}{|l|l|}
\hline
 Example 1 & Illustrate the basic idea of code construction. \\ \hline
 Example 2 & Coding beats routing. \\ \hline
 Example 3 & Class-I Vs. Class-II. \\ \hline
 Example 4 &  $(n_1, 2)$ Vs. $(2, n_2)$. \\ \hline
 Example 5 & A hard example where $|\mathcal{A}|=5$. \\ \hline
 Example 6 & A hard example where $|\mathcal{A}|=12$. \\ \hline
\end{tabular}
\caption{A summary of all the examples}
\label{tab:3}
\end{center}
\end{table}

Systematic coding schemes for achieving the Shannon bound are few and far between.  One of the stumbling blocks for us to achieve the Shannon bound is our lack of a  complete understanding of information inequalities.
By studying the tightness of the Shannon bounds and their associated coding schemes on some networks with simple topologies, we may gain new insights that may help us to further understand $H(K)/H(M)$ for general networks.
For   general Class-I/II ($n_1$, $n_2$) networks, we have numerically computed all three bounds for all the wiretap patterns for the  $(n_1=3, n_2=3)$ case. We also list some wiretap patterns in which the determination of the optimal $H(K)/H(M)$ is challenging.
The examples stated in Sec. \ref{hdexam}  indicate that the determination of the entropic region of $6$ linear vector spaces $\tilde{L}_6$ cannot be sidestepped, since the solution of some complicated wiretap patterns may depend on $\tilde{L}_6$ (see Remark \ref{rmk:1}).
Optimistically speaking, we may conjecture that the Shannon bound is tight in Class-I/II ($n_1$, $n_2$) networks  and the bounds  can be achieved by  linear coding  schemes. Resolving this conjecture is of great interest in information theory. To do so requires us to enhance both the mathematical theory behind information-theoretic security and to aid in   numerical investigations, more sophisticated software will be useful.



 \section*{Acknowledgements}
 The authors are supported by NUS grants R-263-000-A98-750/133 and an NUS Young Investigator Award R-263-000-B37-133.

 This work was also partially funded by a grant from the University Grants Committee of the Hong Kong Special Administrative Region (Project No.\ AoE/E-02/08) and  Shenzhen Key Laboratory of Network Coding Key Technology and Application, Shenzhen, China (ZSDY20120619151314964).

\end{document}